\documentclass[aps,prb,twocolumn,superscriptaddress]{revtex4-2}

\usepackage{amsmath}
\usepackage{amsfonts}
\usepackage{float}
\usepackage{graphicx}	

\usepackage[colorlinks=true]{hyperref}	

\begin{document}

	\title{Robust Majorana bound states in magnetic topological insulator nanoribbons \\ with fragile chiral edge channels}
	
	\author{Declan Burke}
	\affiliation{Blackett Laboratory, Imperial College London, South Kensington Campus, London SW7 2AZ, United Kingdom}
	\author{Dennis Heffels}
	\affiliation{Peter Grünberg Institute 9, Forschungszentrum Jülich, 52425 Jülich, Germany}
    \affiliation{JARA-Fundamentals of Future Information Technology, Jülich-Aachen Research Alliance, Forschungszentrum Jülich and RWTH Aachen University, 52425 Jülich, Germany}
	\author{Kristof Moors}
	\email[]{k.moors@fz-juelich.de}
	\affiliation{Peter Grünberg Institute 9, Forschungszentrum Jülich, 52425 Jülich, Germany}
    \affiliation{JARA-Fundamentals of Future Information Technology, Jülich-Aachen Research Alliance, Forschungszentrum Jülich and RWTH Aachen University, 52425 Jülich, Germany}
	\author{Peter Sch\"uffelgen}
	\affiliation{Peter Grünberg Institute 9, Forschungszentrum Jülich, 52425 Jülich, Germany}
    \affiliation{JARA-Fundamentals of Future Information Technology, Jülich-Aachen Research Alliance, Forschungszentrum Jülich and RWTH Aachen University, 52425 Jülich, Germany}
	\author{Detlev Gr\"utzmacher}
	\affiliation{Peter Grünberg Institute 9, Forschungszentrum Jülich, 52425 Jülich, Germany}
    \affiliation{JARA-Fundamentals of Future Information Technology, Jülich-Aachen Research Alliance, Forschungszentrum Jülich and RWTH Aachen University, 52425 Jülich, Germany}
    \affiliation{JARA-FIT Institute: Green IT, Jülich-Aachen Research Alliance, Forschungszentrum Jülich and RWTH Aachen University, 52425 Jülich, Germany}
	\author{Malcolm R.\ Connolly}
	\email{m.connolly@imperial.ac.uk}
	\affiliation{Blackett Laboratory, Imperial College London, South Kensington Campus, London SW7 2AZ, United Kingdom}
	
	\date{\today}
	
	\begin{abstract}
		Magnetic topological insulators in the quantum anomalous Hall regime host ballistic chiral edge channels. When proximitized by an $s$-wave superconductor, these edge states offer the potential for realizing topological superconductivity and Majorana bound states without the detrimental effect of large externally-applied magnetic fields on superconductivity. Realizing well-separated unpaired Majorana bound states requires magnetic topological insulator ribbons with a width of the order of the transverse extent of the edge state, however, which is expected to bring the required ribbon width down to around $100\,$nm. In this regime, it is known to be extremely difficult to retain the ballistic nature of chiral edge channels and realize a quantized Hall conductance. In this paper, we study the impact of disorder in such magnetic topological insulator nanoribbons and compare the fragility of ballistic chiral edge channels with the stability of Majorana bound states when the ribbon is covered by a superconducting film. We find that the Majorana bound states exhibit greater robustness against disorder than the underlying chiral edge channels. 
	\end{abstract}
	
	
	\maketitle

\section{Introduction}
The realization of Majorana bound states (MBS) and non-Abelian particle exchange statistics would be a watershed moment in condensed matter physics and usher in the era of intrinsically fault-tolerant quantum computing~\cite{Freedman2003, Alicea2011, Aasen2016}. The leading platforms for hosting MBS comprise $s$-wave superconductors (SCs) such as Nb and Al coupled with nanowires of strong spin-orbit III-V semiconductors (InAs, InSb) ~\cite{lutchyn2010majorana}, or V-VI and III-VI topological insulators (TIs) (BiSbTe, HgTe)~\cite{fu2008superconducting, Cook2011}. While electron mobility in III-V nanowires suggests disorder can hamper reliable MBS formation~\cite{Ahn2021}, TI nanoribbons host topological surface states (TSSs) whose innate immunity to bulk disorder could boost MBS stability~\cite{Cook2011, Cook2012, Breunig2022, Heffels2023}. Chiral edge channels (CECs) that form when TIs incorporate magnetic atoms leading to the quantum anomalous Hall (QAH) effect~\cite{Yu2010, Chang2013, Tokura2019, Deng2020} are particularly attractive for obtaining ballistic transport without external magnetic fields. Delocalized chiral Majorana edge modes in millimeter-scale QAH devices~\cite{Qi2010, He2017} have proved challenging to proximitize and detect~\cite{Kayyalha2020}. Pristine quasi-one-dimensional (quasi-1D) magnetic TI nanoribbons (MTINRs) with out-of-plane ferromagnetism and proximity-induced superconducting pairing~\cite{Zeng2018, Chen2018} are amenable to detection with conventional tunneling spectroscopy~\cite{atanov2024proximity} and further incorporation within advanced qubit control architectures~\cite{Manousakis2017, Schueffelgen2019, Schmitt2022}. Understanding the precise role of disorder, however, is crucial. In this paper, we study how disorder in MTINRs influences the emergence and robustness of quasi-1D topological superconductivity and the stability of MBSs.

\begin{figure*}[htb]
  \centering
  \includegraphics[width=0.99\linewidth]{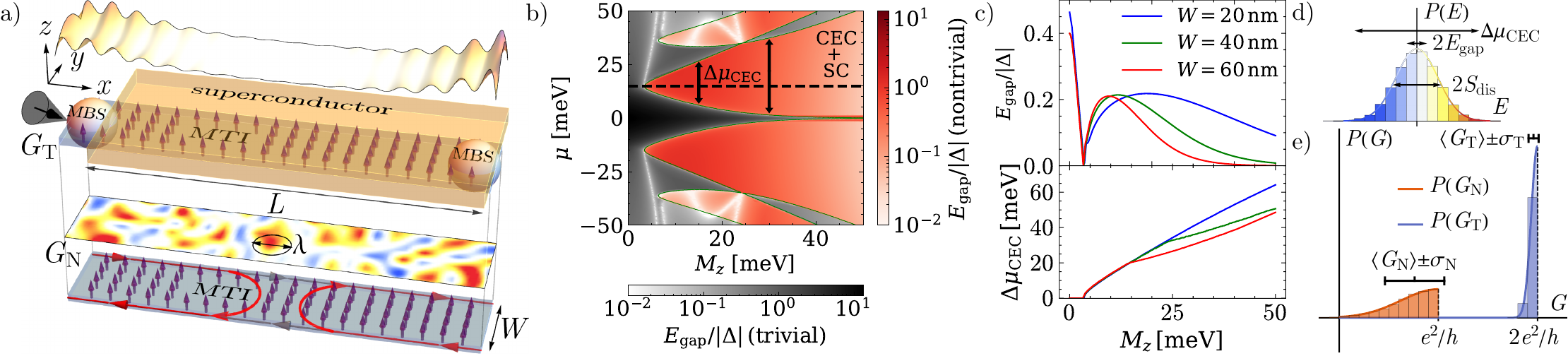}
  \caption{
    (a) Schematic of the MTINR setup: (top) proximitized MTINR with MBSs at opposite ends (MBS wave function density shown above), (middle) a spatially correlated electrostatic disorder profile, (bottom) inter-edge scattering of CECs.
    (b) The spectral gap in a proximitized $40\,\textnormal{nm}$-wide MTINR as a function of magnetization $M_z$ and chemical potential $\mu$, with trivial (nontrivial) regions indicated by a gray (red) color scale, delineated with green boundaries.
    (c) The spectral gap and chemical potential window of the proximitized CEC phase as a function of magnetization $M_z$ for different widths of the MTINR.
    (d) A Gaussian distribution for fluctuations of the MTINR spectrum due to electrostatic disorder, with relevant energy scales (proximity-induced spectral gap $E_\mathrm{gap}$ and extent of the CEC phase $\Delta \mu_\mathrm{CEC}$) indicated.
    (e) Distributions of normal-state and zero-bias tunneling conductance over an ensemble of disordered MTINRs.
  }
  \label{fig:fig1}
\end{figure*} 

\section{Setup}

Figure~\ref{fig:fig1}a shows a schematic of the system under consideration: a MTINR with width $W$ that is covered by a superconducting film over a length $L$. To enable tunneling spectroscopy we include a $100\,\textnormal{nm}$-long normal region at one end of the MTINR. When the MTINR is in the single-channel regime with a single pair of counterpropagating CECs (referred to as the CEC phase and corresponding to the QAH regime for wide MTI ribbons) that get coupled by a proximity-induced superconducting pairing potential $\Delta$, a nontrivial gap opens with unpaired MBSs at opposite ends of the proximitized section~\cite{Chen2018}. In order to open a proximity-induced spectral gap $E_\mathrm{gap}$, the wave functions of the counterpropagating CECs on opposite edges of the ribbon must overlap across the SC-MTI interface (i.e., the top surface in our setup). Unlike nonchiral III-V and undoped TIs, where spatial co-location of particle and hole states makes it straightforward to obtain a spectral gap $E_\mathrm{gap} \sim |\Delta|$, the MTINR width must be comparable to the characteristic width of the CEC $\xi_{\mathrm{CEC}} \approx \hbar v_{\mathrm{D}} / E_{\mathrm{QAH}} \sim 10 \textnormal{--} 100\,\text{nm}$ for $E_{\mathrm{QAH}} \sim 3\textnormal{--}30 \,\mathrm{meV}$~\cite{Zhou2023}, where $E_\mathrm{QAH}$ is the QAH gap that results from the interplay of the magnetization and hybridization of top and bottom surface states~\cite{Yu2010}, and $v_{\mathrm{D}} \sim 4.5\times10^5\,\textnormal{m/s}$ is the Dirac velocity of the (M)TI surface-state spectrum. If the ribbon width significantly exceeds this $\sim10$--$100\,$nm length scale, the overlap and spectral gap will be exponentially suppressed. Thus, the central question for the MBS platform based on MTINRs in the CEC regime is whether sufficiently narrow MTINRs can be realized with enough CEC overlap such that a sizable proximity-induced gap can be opened, while retaining the desired resilience against disorder.

In this paper, we assume that the MTINR forms a single magnetic domain, i.e., a domain without sign changes of the out-of-plane magnetization, enabling CEC formation over a chemical potential window $\Delta\mu_\mathrm{CEC}$ that is proportional to the magnetization strength. We consider a spatially correlated electrostatic disorder landscape directly observed using local tunneling in single-crystal flakes of Cr-doped (Bi$_{1-x}$Sb$_x$)$_2$Te$_3$ exfoliated in UHV~\cite{Chong2020}.

\section{Model}
For a detailed simulation of proximitized MTINRs with disorder, we consider the following modeling approach. We start with the MTI thin-film model Hamiltonian~\cite{Yu2010, Wang2015}:
\begin{equation}
\begin{split}
    H_\text{MTI}(\mathbf{k}) &= \hbar v_\mathrm{D} (k_y \sigma_x - k_x \sigma_y) \rho_z \\
    &\hphantom{=} + [m_0 + m_1 (k_x^2 + k_y^2)] \rho_x + M_z \sigma_z,
\end{split}
\end{equation}
with a two-dimensional wave vector $\mathbf{k} \equiv (k_x, k_y)$, $\sigma_{x,y,z}$ ($\rho_{x,y,z}$) Pauli matrices acting on the spin up-down (top-bottom surface) two-level subspace, and out-of-plane magnetization strength $M_z$.
We consider the following parameters~\cite{Chen2017, Chen2018}: $\hbar v_\mathrm{D} = 3\,\text{eV}\,\text{\r{A}}$, $m_0 = -5\,\text{meV}$, and $m_1 = 15\,\text{meV}\,\text{\r{A}}^2$.
Proximity-induced superconductivity is treated with the Bogoliubov-de Gennes (BdG) formalism and an induced superconducting pairing potential $\Delta$ on the top surface of the MTI thin film or ribbon, yielding the following BdG Hamiltonian:
\begin{equation}
    H_\text{BdG}(\mathbf{k}) =
    \begin{pmatrix}
        H_\text{MTI}(\mathbf{k}) - \mu & -i\sigma_y (1 + \rho_z) \Delta/2 \\
        i \sigma_y (1 + \rho_z) \Delta^\ast/2 & \mu - H_\text{MTI}^\ast(-\mathbf{k})
    \end{pmatrix},
\end{equation}
with the chemical potential $\mu$ and spinor $(\Psi, \Psi^\dagger)$ as the basis vector, with $\Psi = (\mid \uparrow \text{t} \rangle, \mid \uparrow \text{b} \rangle, \mid \downarrow \text{t} \rangle, \mid \downarrow \text{b} \rangle)$ and t (b) referring to the top (bottom) MTI surface.
Disorder is treated by adding a disorder potential $S_\mathrm{dis} (\mathbf{r})$ to the Hamiltonian with a Gaussian random-field profile (see Figs.~\ref{fig:fig1}a and \ref{fig:fig1}d):
\begin{equation}
    \langle S_\mathrm{dis}(\mathbf{r}) S_\mathrm{dis} (\mathbf{r}') \rangle = S_\mathrm{dis}^2 \exp[-(\mathbf{r} - \mathbf{r}')^2/(2\lambda^2)],
\end{equation}
with disorder strength $S_\mathrm{dis}$ and spatial correlation length $\lambda$. Note that we focus on electrostatic (nonmagnetic) disorder here.
The BdG Hamiltonian is discretized to obtain a tight-binding model on an artificial square lattice with the lattice constant equal to $1\,\text{nm}$.
The simulations are performed with Python-based simulation package \textsc{kwant}~\cite{Groth2014}, with parallelized sparse direct solver \textsc{mumps}~\cite{Amestoy2001} and \textsc{adaptive}~\cite{Nijholt2019} for efficient parameter sampling.

\section{Spectral gap}
In Figure~\ref{fig:fig1}b, we present the spectral gap of a $40\,\textnormal{nm}$-wide MTINR superposed with the (trivial or nontrivial) $\mathbb{Z}_2$ topological invariant for a Majorana quantum wire~\cite{Kitaev2001} as a function of chemical potential $\mu$ and magnetization $M_z$ (typical magnetization values are $|M_z| = 10-100\,\textnormal{meV}$ for intrinsic or magnetically doped MTIs~\cite{Lee2015, Rienks2019, Chong2020, Liu2022}). The trivial and nontrivial regions are separated by gapless boundaries, with additional gapless boundaries appearing within (non)trivial regions with multiple (sub)bands~\cite{Chen2018}. This has been seen in spectra of similar systems~\cite{Nijholt2016, Heffels2023}, and is related to the arrangement of multiple bands in reciprocal space. Here, we focus on the chemical potential window with a single band crossing (i.e., a CEC at zero energy), however, which yields a topologically nontrivial phase and is denoted as $\Delta\mu_\mathrm{CEC}$ (indicated for $|M_z| = 15\,\textnormal{meV}$ and $30\,\textnormal{meV}$ in Fig.~\ref{fig:fig1}b). This window is proportional to $|M_z|$ and thus easily an order of magnitude larger than the typical size of the proximity-induced spectral gap $E_{\mathrm{gap}} \lesssim 0.2 |\Delta|$ (see Fig.~\ref{fig:fig1}c). To examine the interplay between disorder and MBSs (CECs) in (non)proximitized MTINRs and benchmark their robustness, we calculate both the mean and standard deviation of the tunneling (normal-state) conductance distributions, as schematically depicted in Fig.~\ref{fig:fig1}e.

\begin{figure}[tb]
  \centering
  \includegraphics[width=.99\linewidth]{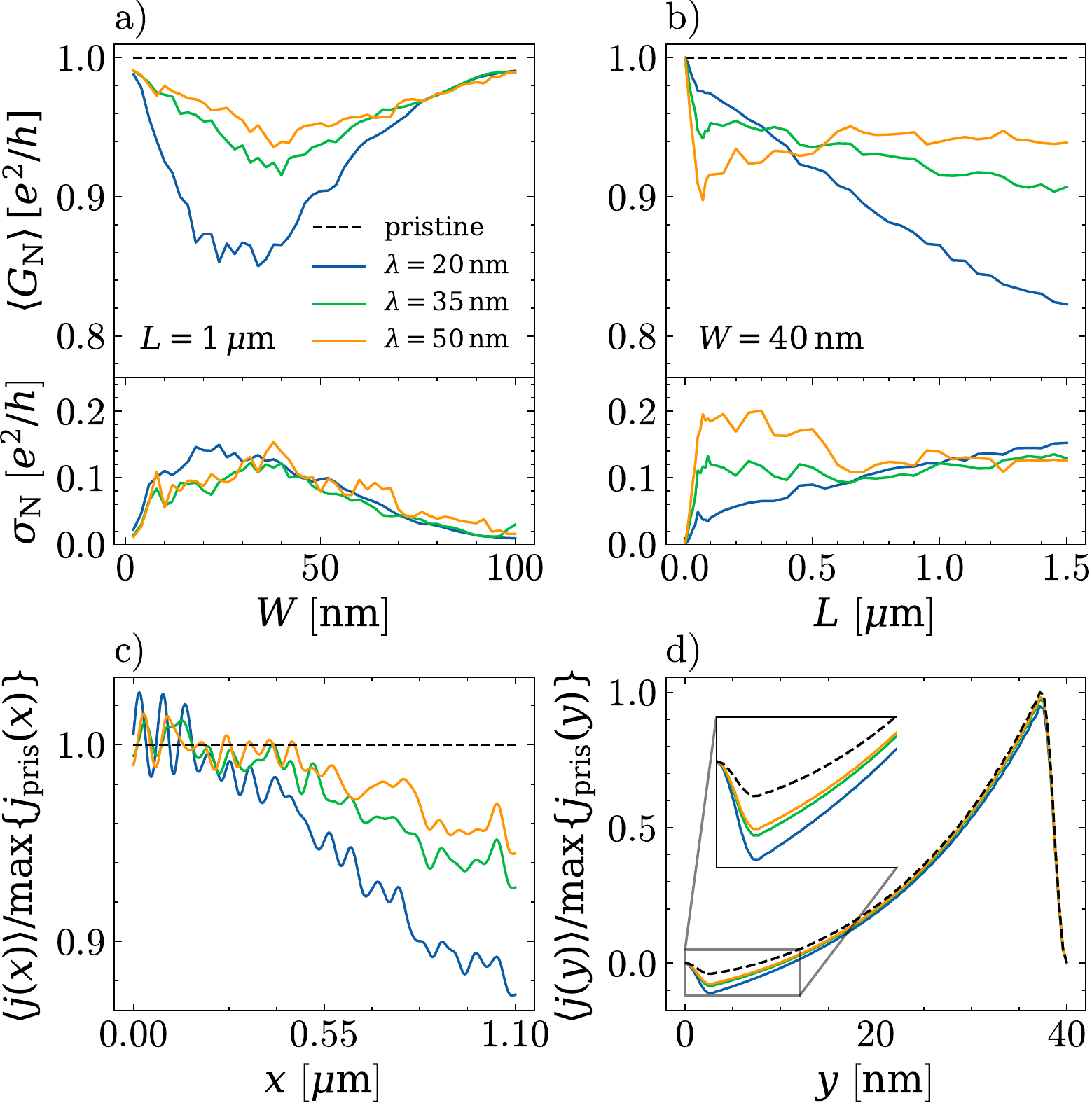}
  \caption{
    [(a),(b)] The normal-state two-terminal conductance (top) mean and (bottom) standard deviation of 200 MTINRs with $M_z = 15\,$meV in the single-channel regime as a function of (a) width and (b) length, considering disorder strength $S_\mathrm{dis} = 3\,\textnormal{meV}$ and different spatial correlation lengths.
    [(c),(d)] The local current density along the transport direction of a forward-propagating CEC in a $40\,\textnormal{nm}$-wide and $1\,\mu\textnormal{m}$-long MTINR with the same disorder statistics and averaging as in (a) and (b) is shown as a function of (c) the coordinate along the MTINR (integrated over the transverse coordinate $y$, see Fig.~\ref{fig:fig1}a) and of (d) the transverse coordinate (integrated over the transport direction).} 
  \label{fig:fig2}
\end{figure}

\section{Conductance}
\subsection{Normal-state conductance}
The normal-state conductance of a MTINR in the CEC phase is shown as a function of MTINR dimensions for different disorder correlation lengths in Fig.~\ref{fig:fig2}. For each ribbon width, the chemical potential is fixed to a value for which the topological region of interest is reached with the smallest possible value of $M_z$ (see horizontal linecut in Fig.~\ref{fig:fig1}b). This guarantees that the chemical potential is well separated from the phase boundaries of the topological region (which is essential for the robustness against disorder fluctuations) for all values of $M_z$. For a $1\,\mu\textnormal{m}$-long MTINR, we observe near-perfect transmission for small and large widths, independent of the disorder correlation length (see Fig.~\ref{fig:fig2}a). For large widths, this is expected due to CEC separation, while for small widths the system enters a quasi-1D Klein tunneling regime. For intermediate widths $W \sim \lambda$, i.e., comparable to the disorder correlation length, the disorder-induced backscattering probability becomes maximal and, correspondingly, we observe a minimum in the conductance. Note that the spatial correlation statistics of the disorder profile become irrelevant when the ribbon width significantly exceeds the spatial correlation length. Interestingly, the conductance has an approximately linear decrease as a function of length when the length significantly exceeds the disorder correlation length and the decrease is more pronounced for a shorter correlation length (see Fig.~\ref{fig:fig2}b), which indicates a conventional ohmic transport regime. This is corroborated by a linear decrease of the edge-current density along the length of the ribbon, which is revealed by removing quantum interference-induced current-density fluctuations through averaging over multiple samples (see Fig.~\ref{fig:fig2}c). As we are considering standard Landauer-B\"uttiker theory, with the two-terminal conductance $G=\tau e^2/h$ being related to the CEC transmission probability $\tau$, our results indicate that disorder introduces a homogeneous distribution of inter-edge scattering sites, which yields quasi-diffusive (ohmic) CEC transport. Due to MTINR confinement, the CECs hybridize and become partially co-located on the same edge, and disorder can easily furnish direct backscattering to the counterpropagating CEC. This can be clearly seen in Fig.~\ref{fig:fig2}d, where a reduction of conductance is manifested as a more pronounced negative current density peak near the edge of the counterpropagating CEC. Hence, CECs in MTINRs are generally fragile with respect to disorder, which impedes the realization of a quantized two-terminal conductance across the ribbon (or quantized Hall conductance in a Hall bar setup).

\begin{figure}[tb]
  \centering
  \includegraphics[width=.99\linewidth]{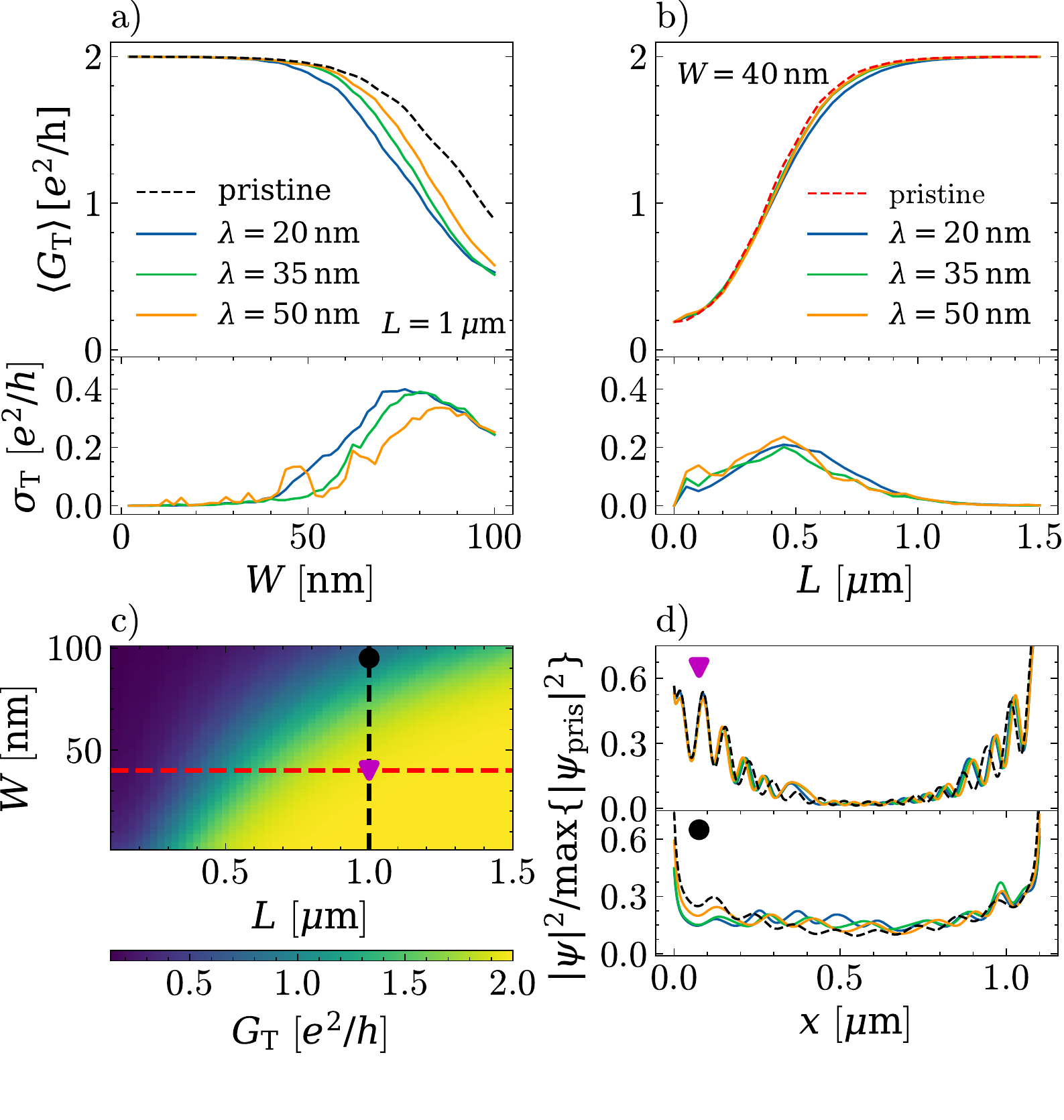}
  \caption{[(a),(b)] The tunneling conductance (top) mean and (bottom) standard     deviation as a function of (a) width and (b) length of 200 proximitized MTINRs   with identical magnetization, dimensions, and disorder statistics as considered in Fig.~\ref{fig:fig2}.
    (c) The tunneling conductance $G_\mathrm{T}$ of a pristine MTINR as a function of ribbon length $L$ and width $W$.
    (d) The MBS wave function density of $1\,\mu\textnormal{m}$-long proximitized MTINRs with different widths [indicated in (c)] in the pristine ($S_\mathrm{dis} = 0\,\textnormal{meV}$) and disordered ($S_\mathrm{dis} = 3\,\textnormal{meV}$) case, with different spatial correlation lengths [legends in (a),(b)].}
  \label{fig:fig3}
\end{figure}    

\subsection{Tunneling conductance}
Next, we consider the tunneling conductance (of a metallic tunneling probe, schematically depicted by the tip in Fig.~\ref{fig:fig1}a) at zero bias (see Fig.~\ref{fig:fig3}) for similar MTINRs as before, but with a proximity-induced pairing potential $\Delta$ on the top surface. As the MBS localization length $\xi_\mathrm{MBS}$ is inversely proportional to the pairing potential ($\xi_\mathrm{MBS} \propto 1 / |\Delta|$), a smaller value for $\Delta$ can be compensated by increasing the MTINR length for obtaining well-separated MBSs ($L \gg \xi_\mathrm{MBS}$). Hence, the results should not change qualitatively under a rescaling $L \propto 1/|\Delta|$ such that the ratio $L/\xi_\mathrm{MBS}$ remains unchanged, which is why we consider a rather large top-surface pairing potential ($|\Delta| = 5\,\textnormal{meV}$) to reduce the required system size and, correspondingly, the computational burden. Despite the fragility of the CEC in normal transport, the zero-bias tunneling conductance, i.e., the zero-bias peak (ZBP), is remarkably well quantized at $2e^2/h$ for small ribbon widths, as long as the MTINR is sufficiently long (see Figs.~\ref{fig:fig3}a and \ref{fig:fig3}b). A smooth reduction of the ZBP appears when the ribbon width increases and becomes comparable to the transverse CEC size $\xi_\mathrm{CEC}$, or when the ribbon length decreases and becomes comparable to the MBS localization length $\xi_\mathrm{MBS} \approx \hbar v_\mathrm{D} / E_\mathrm{gap}$ ($\xi_\mathrm{MBS} \approx 0.3\,\mu\textnormal{m}$ for $E_\mathrm{gap} \approx 1\,\textnormal{meV}$). Overall, the robustness of the ZBP is dictated by the width and length of the MTINR (see Fig.~\ref{fig:fig3}c), which control the inter-edge CEC overlap (across the ribbon width) and the inter-end MBS separation (along the ribbon length), respectively. Note that the required length scales exponentially with the ribbon width for large widths, due to the exponential suppression of the CEC overlap and proximity-induced spectral gap, $\propto |\Delta| \exp(- W/\xi_\mathrm{CEC})$ (see the Appendix for details).
Disorder barely affects the appearance of well-separated MBSs (see Fig.~\ref{fig:fig3}d) or the resulting tunneling conductance. The ZBP is only significantly affected by disorder when the quantized ZBP of the pristine MTINR is already in a breakdown regime. Hence, while the formation of well-separated MBSs requires long ($L \gg \xi_\mathrm{MBS}$) and narrow ($W \lesssim \xi_\mathrm{CEC}$) MTINR dimensions for which quasi-diffusive CEC transport is expected, the MBS formation itself is robust with respect to disorder.

\subsection{Comparison}
\begin{figure}[tb]
  \centering
  \includegraphics[width=0.99\linewidth]{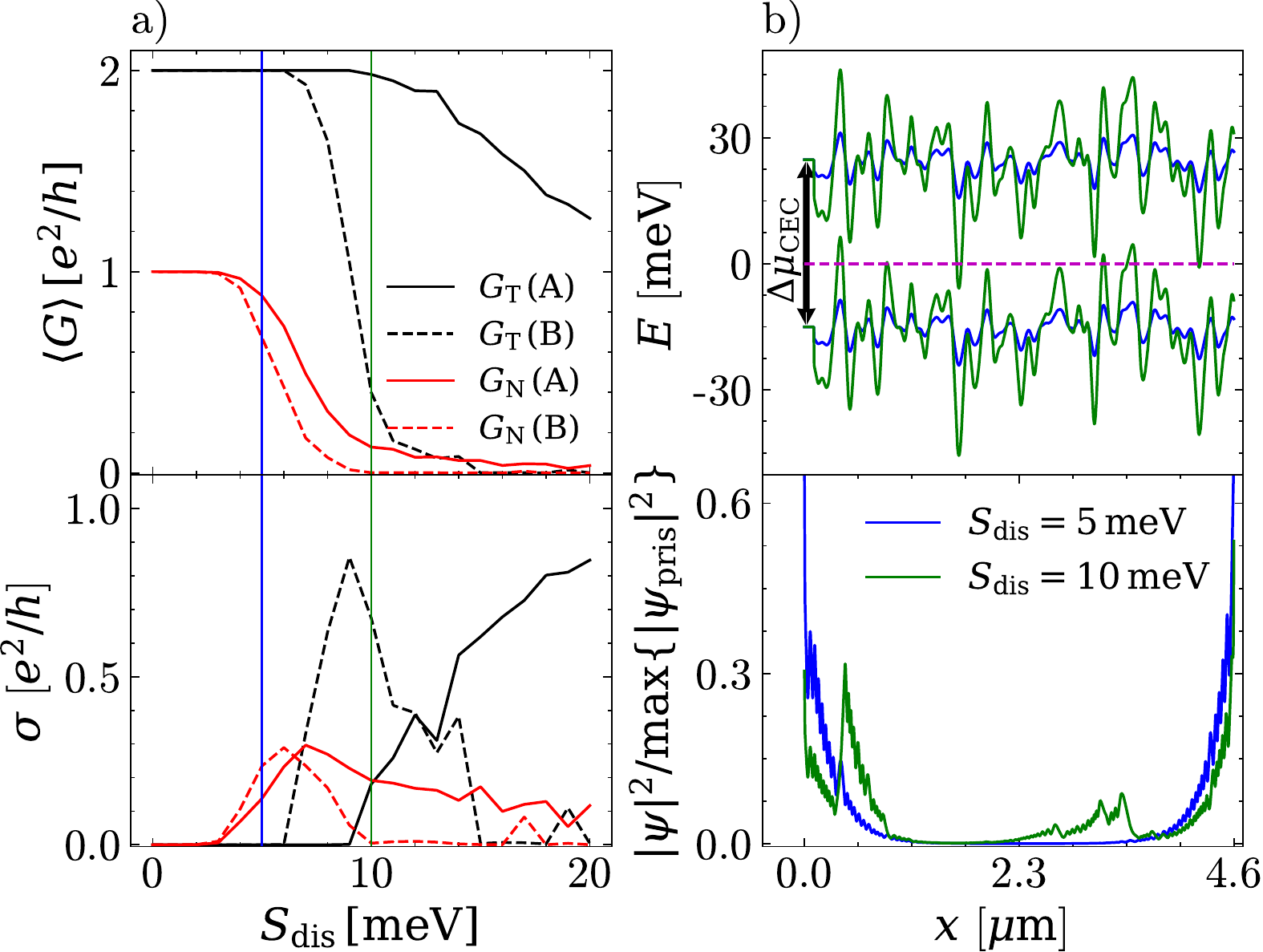}
  \caption{
    (a) The normal-state and tunneling conductance (top) mean and (bottom) standard deviation of 100 MTINRs with $M_z = 30\,\textnormal{meV}$, $W = 30\,\textnormal{nm}$, (A) $L = 4.5\,\mu$m and $|\Delta| = 5\,$meV, or (B) $L = 15\,\mu\textnormal{m}$ and $|\Delta| = 1.5\,\textnormal{meV}$, as a function of disorder strength with $\lambda = 35\,\textnormal{nm}$.
    (b) The (top) local extent of the CEC phase around the chemical potential (horizontal dashed line) and (bottom) wave function density of the two lowest-energy solutions along the length of a MTINR with chemical potential in the single-channel regime (similar to horizontal linecut in Fig.~\ref{fig:fig1}b), considering different disorder strengths [indicated in (a)]. The wave function density is presented relative to the maximum density of the solution of a pristine MTINR (not shown, but nearly indistinguishable from the density with $S_\mathrm{dis} = 5\,$meV).
  } 
  \label{fig:fig4}
\end{figure}

The normal-state and tunneling conductances of a MTINR with fixed dimensions are presented as a function of disorder strength in Fig.~\ref{fig:fig4}. We can see clearly that the breakdown of the quantized ZBP is decoupled from the breakdown of perfect CEC transmission (Fig.~\ref{fig:fig4}a). As discussed above, CEC transmission in MTINRs is critically dependent on inter-edge backscattering due to disorder, which can easily be induced for narrow long MTINRs with relatively weak disorder. For retaining a quantized ZBP, however, it is sufficient that the MTINR remains in the topologically nontrivial phase with a proximitized single (spinless) channel. This requires that the disorder strength is small compared to the CEC window $\Delta\mu_\mathrm{CEC}$ around the chemical potential (see Fig.~\ref{fig:fig4}b).

We also examine the scaling relation between the pairing potential and the ribbon length by considering MTINRs with different values for proximity-induced pairing $\Delta$ and ribbon length $L$ with the same product $|\Delta| L$ and identical disorder characteristics. While the ZBP breaks down when the disorder becomes comparable to the MBS localization length in both cases, there is a notable difference. The MTINR with smaller pairing and larger ribbon length appears to be less robust. This can be understood by considering the detailed statistics of disorder (Fig.~\ref{fig:fig1}d), which are responsible for locally driving the system out of the topologically nontrivial phase. For a long MTINR with weak superconducting pairing, the probability of having local spectral fluctuations that significantly exceed the standard deviation (i.e., the disorder strength) in the sample is larger than for a shorter MTINR with stronger pairing. Such fluctuations nucleate additional low-energy states (MBS pairs) throughout the ribbon that hybridize with the end-localized MBSs (see how the two lowest-energy solutions in Fig.~\ref{fig:fig4}b feature density peaks in the MTINR interior in the case of strong disorder) and push the ZBP breakdown to smaller disorder strengths. Hence, relatively short and strongly proximitized MTINRs are favored for realizing a quantized ZBP, as compared to relatively long and weakly proximitized MTINRs with similar $|\Delta| L$, to minimize outliers of the spectral fluctuations.

\section{Conclusion}
In conclusion, we find that long $\sim$10--100\,nm-wide magnetic topological insulator ribbons are required to realize a topologically nontrivial proximity-induced superconducting gap with well-separated Majorana bound states on opposite ends of the ribbon. These ribbon dimensions render chiral-edge-channel transmission fragile with respect to disorder and a quasi-diffusive transport regime is to be expected in normal transport due to direct backscattering. However, this fragility does not impede the formation of well-separated Majorana bound states with a quantized zero-bias peak in tunneling spectroscopy that remains robust against electrostatic disorder up to an energy scale that is proportional to the magnetization strength.
While considerable challenges must be overcome to fabricate high-quality narrow magnetic topological insulator ribbons in the single-channel regime, the resilience against disorder makes this a very promising platform for realizing Majorana bound states.


\begin{acknowledgments}
We thank Thomas L.\ Schmidt and Llorenç Serra for useful discussions. D.H.\ and K.M.\ acknowledge the financial support by the Bavarian Ministry of Economic Affairs, Regional Development and Energy within Bavaria’s High-Tech Agenda Project ``Bausteine für das Quantencomputing auf Basis topologischer Materialien mit experimentellen und theoretischen Ansätzen'' (Grant Allocation No.\ 07 02/686 58/1/21 1/22 2/23). M.R.C. received funding from EPSRC (EP/L020963/1) and the President's Excellence Fund for Frontier Research. This work is supported by the QuantERA grant MAGMA and by the German Research Foundation under Grant No.\ 491798118. P.S.\ acknowledges financial support by the German Federal Ministry of Education and Research (BMBF) via the Quantum Futur project ``MajoranaChips'' (Grant No.\ 13N15264) within the funding program Photonic Research Germany.
\end{acknowledgments}

\appendix
\renewcommand\thefigure{A\arabic{figure}}
\setcounter{figure}{0}

\section*{Appendix: Length and width scaling of the ribbon}

\begin{figure}[htb]
  \centering
  \includegraphics[width=0.99\linewidth]{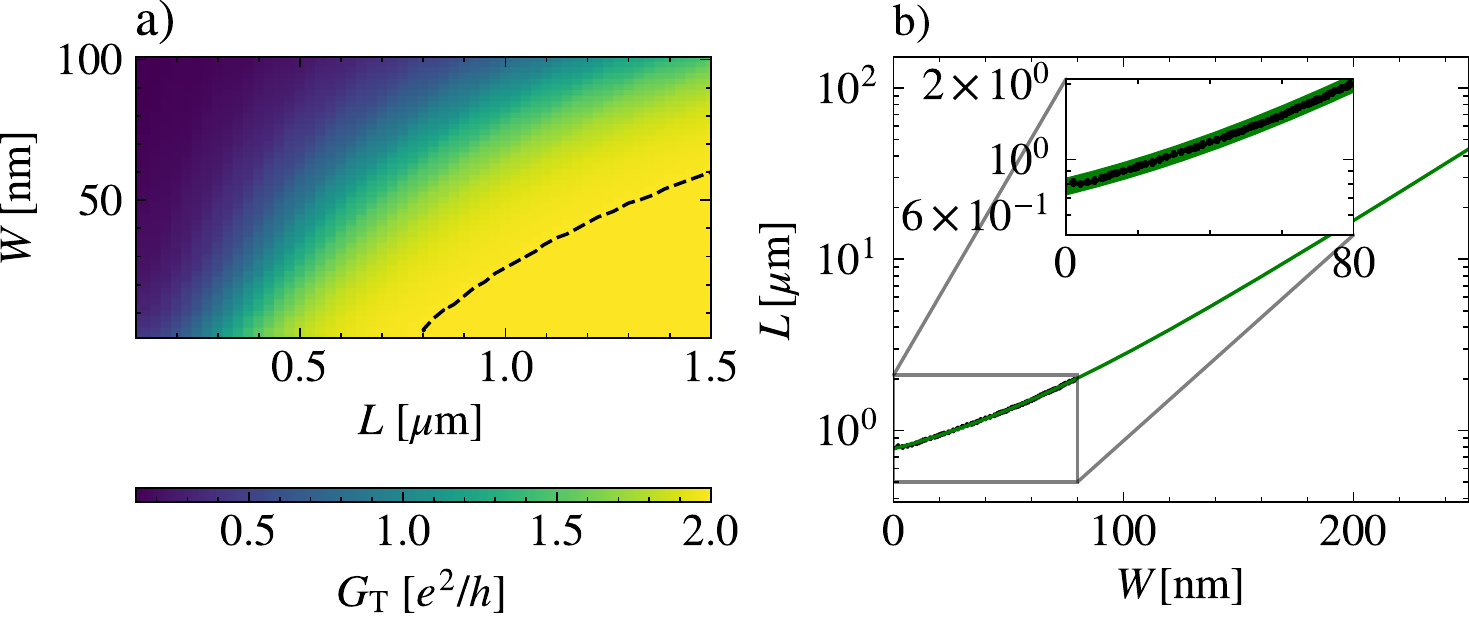}
  \caption{(a) The tunneling conductance shown in Fig.~\ref{fig:fig3}a, with contour line of constant tunneling conductance $G_\mathrm{T} = 1.99\, e^{2}/h$. (b) The contour line (shown as dots) and the exponential fitting function, which is extrapolated to larger ribbon widths and lengths.}
  \label{fig:A1}
\end{figure}

While a proximitized magnetic topological insulator nanoribbon should be sufficiently narrow and long to obtain well-separated MBSs with a nearly quantized zero-bias conductance peak (see Fig. 3), there is still the freedom to choose different aspect ratios. Due to the exponential suppression of the CEC overlap and proximity-induced spectral gap, however, an exponential increase of the required ribbon length is expected, when increasing the width significantly above the CEC width, in order to maintain equally well-separated MBSs. We verify this scaling relation for the aspect ratio by fitting a contour of constant tunneling conductance, $G_\mathrm{T} = 1.99\, e^2/h$, denoted by a dashed line in Fig.~\ref{fig:A1}a, with the following fitting function: $W_\mathrm{f} + C_\mathrm{f} \, \mathrm{exp}(W/\xi_\mathrm{f})$ (see Fig.~\ref{fig:A1}b). The function fits very well to the contour with the following parameters: $W_\mathrm{f} = 0.39\,\mu\textnormal{m}$, $C_\mathrm{f} = 0.025\,\mu\textnormal{m}$, and $\xi_\mathrm{f} = 50\,\textnormal{nm}$. The exponential has a characteristic length scale $\xi_\mathrm{f}=50\,\textnormal{nm}$, which is comparable to the CEC width, $\xi_\mathrm{CEC} = \hbar v_\mathrm{D}/E_\mathrm{QAH} = 30\,\textnormal{nm}$, for the model parameters under consideration. Because of this scaling relation, a $\sim$10--100\,nm ribbon width will be necessary to avoid impractically long ribbons from an experimental point of view. Furthermore, longer ribbons are more likely to fluctuate locally out of the topological regime due to outliers in the disorder profile, and should therefore also be avoided.

\bibliographystyle{apsrev4-2}
\bibliography{references}

\end{document}